\documentstyle[12pt,aps,epsf]{revtex}
\def\er#1#2{\relax\ifmmode{}^{+#1}_{-#2}\else$^{+#1}_{-#2}$\fi}
\newcommand{\be}{\begin{equation}}
\newcommand{\bea}{\begin{eqnarray}}
\newcommand{\ee}{\end{equation}}
\newcommand{\eea}{\end{eqnarray}}

\newcommand{\krig}[1]{\stackrel{\circ}{#1}}
\def\({\Big(}
\def\){\Big)}

\def\slashchar#1{\setbox0=\hbox{$#1$}
   \dimen0=\wd0 \setbox1=\hbox{/} \dimen1=\wd1
   \ifdim\dimen0>\dimen1 \rlap{\hbox to \dimen0{\hfil/\hfil}} #1
   \else  \rlap{\hbox to \dimen1{\hfil$#1$\hfil}} / \fi}

\begin{document}
\tighten
\def\footnoterule{\kern-3pt \hrule width\hsize \kern3pt}

\title{
Improved Unitarized Heavy Baryon Chiral Perturbation Theory
for $\pi N $ Scattering\footnote{Work partially 
supported  by the Junta de Andaluc\'\i a and DGES PB98 1367
and by DGICYT under 
contracts AEN97-1693 and PB98-0782.} 
}
\author{A. G\'omez Nicola$^a$\footnote{email:gomez@eucmax.sim.ucm.es},
J. Nieves$^b$\footnote{email:jmnieves@ugr.es},
J.R. Pel\'aez$^a$\footnote{email:pelaez@eucmax.sim.ucm.es},
 and E. Ruiz Arriola$^b$\footnote{email:earriola@ugr.es}
} 

\address{
{~} \\
$^a$Departamento de F\'{\i}sica Te\'orica\\
Universidad Complutense. 28040 Madrid, Spain.\\
{~} \\
$^b$Departamento de F\'{\i}sica Moderna \\
Universidad de Granada.
E-18071 Granada, Spain
}
\date{\today}
\maketitle

\thispagestyle{empty}

\begin{abstract}
We show how the unitarized description of pion nucleon scattering
within Heavy Baryon Chiral Perturbation Theory can be considerably
improved, by a suitable reordering  of the expansion over the nucleon
mass. Within this framework, the $\Delta$ resonance and its associated
pole can be recovered from the chiral parameters obtained from
low-energy determinations. In addition, we can obtain a good
description of the six $S$ and $P$ wave phase shifts in terms of
chiral parameters with a natural size and compatible with the
Resonance Saturation Hypothesis.
\end{abstract}

\vspace*{1cm}
\centerline{\it PACS: 11.10.St;11.30.Rd; 11.80.Et; 13.75.Lb;
14.40.Cs; 14.40.Aq\\}
\vspace*{1cm}
\centerline{\it Keywords: 
Chiral Perturbation Theory, Unitarity, $\pi N$-Scattering, }
\centerline{\it $\Delta$ Resonance, Heavy baryon, Partial waves.}

\newpage
\setcounter{page}{1}

\section{Introduction}

The modern way to incorporate  chiral symmetry and departures
from it in low energy hadron dynamics, 
is by means of Chiral Perturbation Theory
(ChPT)\cite{gl84}. Being  an effective Lagrangian approach 
all the detailed information on higher energies or underlying
microscopic dynamics is effectively encoded in some low
energy coefficients (LEC) which have to be  determined experimentally. 
For processes involving only pseudoscalar
mesons the expansion parameter is $ p^2/(4\pi F)^2 $
with $p$ their four momentum and $F$ the
weak pion decay constant \cite{gl84}. Of course, this expansion works
better in the threshold region, breaking down 
at higher energies where the violation of unitarity
becomes more and more severe. However, it has been shown that
this applicability region can be extended by means of unitarization
methods, describing remarkably
well meson-meson scattering and its light resonances up to
almost 1.2 GeV \cite{IAM,Nosotros}.
In addition, it has been shown that the 
{\it predicted} unitarized amplitudes can reproduce the threshold region,
and provide definite theoretical central values and error estimates
of the  phase-shifts away from threshold and up to about 1 GeV
\cite{EJ99}.

Baryons can also be included as explicit
degrees of freedom if they are treated as heavy
particles in a covariant framework \cite{IW89}
called Heavy-Baryon Chiral Perturbation Theory (HBChPT)
\cite{JM91,BK92,BKM95}. In this case, the expansion to order 
$N=1,2,3, \dots$
is written in terms of contributions 
of the form $ e^N /( F^{2l}
M^{N+1-2l} ) $, where $l=1, \dots , [(N+1)/2] $, and $M$ is
the baryon mass. The
quantity $e$ is a generic parameter with dimensions of energy built up
in terms of the pseudoscalar momenta and the velocity $v^\mu$ ($ v^2
=1$ ) and off-shellness $ k $ of the baryons. The latter is defined
from $p_B=\krig{M}v +k $, where $p_B$ and $\krig{M}$ 
are the baryon four momentum and its mass 
at lowest order in HBChPT, respectively. 

Within HBChPT $\pi N$ scattering, has been calculated 
up to third order \cite{Mo98,FMS98} so far\footnote{\label{foot} 
After submitting this work the 
HBChPT fourth order result has appeared in the literature
\cite{4orden}.}. In general, it
is found that the HBChPT convergence is not as good as that
of pure ChPT
(see remarks in Ref.~\cite{Mo98} and also below). 
Therefore, any attempt to unitarize this amplitude
based on the increasing smallness of higher order terms,
may {\it formally} reproduce HBChPT series but  will
likely fail {\it numerically} to describe the corresponding phase
shifts even in the threshold region. Hence, any
unitarization method should take into this slow convergence.

Unitarization methods have also been applied to $\pi N$ scattering
in the literature. Already in the seventies Pad\'e approximants
had been used to unitarize simple phenomenological models \cite{Basdevant},
and a more systematic approach based on an effective Lagrangian formalism
was called for. In addition, several relativistic phenomenological models
exist which unitarize tree level amplitudes with the K-matrix method
providing a reasonably good description of $\pi N$ scattering \cite{Kpin}.
More recently, the $\Delta$ and $N^*$ resonances
have also been considered as explicit
degrees of freedom  within HBChPT~\cite{ET97,HH97,MO99}. This
requires the introduction of new parameters into the Chiral
Lagrangian. 

In contrast, the unitarization via
the Inverse Amplitude Method (IAM) \cite{IAM} does not 
introduce new parameters. However, when
applied to $\pi N$ scattering~\cite{GP99}, 
considering the higher orders to be increasingly small,
the LEC turned out to be very different from those found within
HBChPT~\cite{Mo98} since they absorb higher order contributions which are not 
negligible. 
Nevertheless, in ~\cite{GP99} it was also shown that 
the $\Delta(1232)$ can be reproduced with the IAM 
using $O(p^2)$ parameters constrained to lie within the range
of values of Resonance Saturation, although the resulting $O(p^3)$
parameters are still very unnatural. This situation is somewhat 
disappointing, since Pad\'e approximants,
which are similar to the IAM, together with
simple phenomenological models, provided good descriptions
of $\pi N$ scattering.

In this work we show how the $O(p^3)$ HBChPT series can be reordered
and the IAM modified, in  order to i)
implement exact unitarity , ii) comply with  HBChPT at threshold and
iii) describe the $\Delta $ resonance without introducing
additional parameters.  

\section{ Partial Wave Amplitudes}

Customarily, low-energy $\pi N$ scattering is described in terms of
partial waves ( see, for instance Refs~\cite{EW88,Mo98}). We will rely
heavily on the results in  Ref.~\cite{Mo98}, where the third order HBChPT
expressions where first obtained.  However, in that work
the full nucleon mass dependence has been retained, causing 
some order mixing. For our purposes,
and in order to keep track of perturbative unitarity (see below),
we have preferred to further expand the partial waves in terms of 
$1/M$ or $1/F^2$. That is, we only keep the  pure
$O(e/F^2)$ (first order), $O( e^2 /(F^2 M) ) $ (second order) as well
as the $O(e^3 /F^4) $ and $O(e^3/(F^2 M^2))$ (third order) terms. 
One thus gets the following expansion, 
\begin{equation}
 f_{l \, \pm}^\pm  =    f_{l \, \pm}^{(1) \, \pm}
                       +f_{l \, \pm}^{(2) \, \pm}
                       +f_{l \, \pm}^{(3) \, \pm}+ \cdots
\label{eq:fexpansion}
\end{equation}
which thus coincide with the first, second and third orders 
in \cite{FMS98}, where the full nucleon mass dependence was not
retained. However, we also separate these contributions in
\begin{eqnarray}
f_{l \, \pm}^{(1) \, \pm} &=&
{m\over F^2} t_{l \, \pm}^{(1,1) \, \pm} \({\omega \over m}\) \nonumber\\
f_{l \, \pm}^{(2) \, \pm} &=&
{m^2 \over F^2 M } t_{l \, \pm}^{(1,2) \, \pm} \({\omega \over m} \) \nonumber\\
f_{l \, \pm}^{(3) \, \pm} &=&
{m^3 \over F^4 } t_{l \, \pm}^{(3,3) \, \pm} \( {\omega \over m} \)
+ {m^3 \over F^2 M^2 } t_{l \, \pm}^{(1,3) \, \pm} \({\omega\over m} \)
\label{eq:fexpansion2}
\end{eqnarray}
where $t^{(n,m) \, \pm}_{l \, \pm}$  are dimensionless functions of the 
dimensionless variable $\omega/m$, independent of $F,M$ and $m$. The
analytical expressions are too long to be displayed here, but can be
easily obtained starting from Ref.~\cite{Mo98}. 
The convenience of the double superscript notation 
will be explained below; $n+1$ indicates the power 
in $1/F$ and $m$ the total 
order in the HBChPT counting. The unitarity condition $ {\rm Im}
f_{l \, \pm}^{-1} = -q $ becomes in perturbation theory 
\begin{eqnarray}
{\rm Im}\, t_{l \, \pm}^{(1,1) \, \pm} = 
{\rm Im}\, t_{l \, \pm}^{(1,2) \, \pm} &=& 
{\rm Im}\, t_{l \, \pm}^{(1,3) \, \pm} = 0 \\
{\rm Im}\, t_{l \, \pm}^{(3,3) \, \pm} &=& {q\over m}
| t_{l \, \pm}^{(1,1) \, \pm} |^2
\end{eqnarray}
The equivalent of the last equation above
for the $f$ amplitudes is not satisfied exactly if the nucleon
mass dependence is retained through a redefinition of the nucleon
field as in Ref.(~\cite{Mo98}) 
(see, for instance, comments in \cite{FMS98} and 
\cite{GP99}), although, of course, the corrections
are just higher order in HBChPT. These exact relations for the perturbative
contributions will be very convenient later on, and, as we have 
commented above, that is 
the reason why we have preferred to expand the amplitudes 
as in Eq.~(\ref{eq:fexpansion}).

The scattering lengths $ a_{l,\pm}^\pm $ and effective ranges $ b_{l,\pm}^\pm$
are defined by
\begin{equation}
{\rm Re} f_{l,\pm}^\pm = q^{2l} 
\Big( a_{l,\pm}^\pm + q^2 b_{l,\pm}^\pm + \cdots \Big)
\end{equation}
Obviously, the expansion of Eq.~(\ref{eq:fexpansion2}) 
carries over to the threshold parameters, $a_{l,\pm}^\pm $
and $b_{l,\pm}^\pm $, which we will use next to illustrate the 
slow low energy convergence of the HBChPT series.

Throughout this paper we use $F=92.4 \,{\rm MeV}$, $M=938.27 \,{\rm MeV}
$, $m = 139.57 \,{\rm MeV} $ and $g_A = 1.26$.  
Concerning the LEC , there are several determinations: The first
one, which we give as Set I in Table I, was obtained from a fit to the
{\it extrapolated} threshold parameters $a_{0,+}^\pm $, $b_{0,+}^\pm
$, $a_{1,\pm}^\pm $, the Goldberger-Treiman discrepancy and the
nucleon $\sigma-$term ~\cite{Mo98}. Note
that the full nucleon mass dependence was retained. For our purposes
it is thus more convenient a more recent determination\cite{FMS98}
 from a low energy fit to $\pi N$ phase-shift data, although the
authors used a different notation for the chiral
 parameters. The
resulting LEC translated to the $a_i$ and $b_i$
notation (see \cite{EM96,FMS98} ) are given in the Set II of Table I. On the theoretical side, 
there are estimations of the $a_i$ parameters assuming they are
saturated by the exchange of resonances \cite{BKM97}, 
in fairly good agreement with experimental determinations.

\section{Low Energy Convergence} 

In all our following 
considerations we further expand the amplitudes and threshold parameters
as in Eqs.(\ref{eq:fexpansion2}). For the scattering lengths we thus have  
\begin{equation}
a_{2I \, 2J}^l  m^{2l+1} = {m\over F^2} \alpha^{(1,1)} + 
{m^2 \over F^2 M } \alpha^{(1,2)} + {m^3 \over F^2 M^2 } 
\alpha^{(1,3)} + {m^3 \over F^4  } \alpha^{(3,3)} + \cdots 
\end{equation}
and a similar expression for $ b_{2I \,2J}^l m^{2l+3} $. 
The slower convergence of the HBChPT series as 
compared with ChPT was pointed out by the authors of the
first calculations, (see for instance the comments in
\cite{BKM97,Mo98,FMS98}).  For our purposes, we have found instructive
to separate the contributions to the scattering lengths and effective
ranges of the lowest partial waves, which are given in Table II, using
the set II of LEC in Table I. Note that,  in this way,
 the threshold parameters are
predictions, unlike those from set I which uses them as an input.  A
distinctive pattern emerging from this Table is that the contribution
of order $1/(F^2 M^2)$ is always rather small.  Only in some cases,
however, is the contribution of order $1/F^4 $ also small. This is so
in the $P_{33}$ channel in particular, which is the one less likely to
be well described by HBChPT alone due to the low mass of the
$\Delta(1232)$ resonance. 
Hence, it seems
that close to threshold the $1/F^2$ expansion converges faster than
the $1/M$ expansion. Actually, the $1/F^2$ and $1/(F^2 M) $ terms
are comparable. It is clear that any
unitarization method will only be consistent with the HBChPT approach,
if it treats both the first and the second order as equally important.

\section{ Unitarization method for the reordered series}

Our unitarization method assumes that, as suggested by the
perturbative calculation, the chiral expansion in terms of 
$1/F^2 $ converges much faster 
 than the finite nucleon mass $1/M $ corrections. 
Indeed there are some recent theoretical attempts\cite{BL99,Ge99} 
to define a relativistic power counting not requiring the heavy baryon
idea, returning somehow to 
the spirit of older relativistic studies~\cite{gss88}. 
As a matter of fact, we show that the well-known IAM approach applied to 
the reordered HBChPT series generates the $P_{33}$ phase-shift 
satisfactorily, just using the LEC determined at low energies.
In addition, it is possible to improve the overall
 description
of the six $\pi N$ scattering S and P waves, using those very same 
parameters. Finally, if one fits the $\pi N$ phase shifts with this method 
(either constraining the parameters to the Resonance
Saturation Hypothesis or leaving all the parameters unconstrained) 
the resulting parameters have a much more natural size, and the
$\chi^2$ per d.o.f is considerably better than those obtained
with the IAM applied to plain HBChPT \cite{GP99}. 
\begin{table}
\vspace{-0.3cm}
\begin{center}
\begin{tabular}{|c||c|c||c|c|}
\hline 
& M.Mojzis \protect{\cite{Mo98}}
&N.Fettes \emph{et al.} \protect{\cite{FMS98}}& Resonance Saturation fit
&Unconstrained fit\\
& Set {\bf I} & 
Set {\bf II} &  Set {\bf III}&  Set {\bf IV}\\ \hline 
$ a_1$  
& $-$ 2.60 $\pm$  0.03  
& $-$ 2.69 $\pm$  0.4  
& $-$2.065 $\pm$  0.007
& $-$ 1.36 $\pm$ 0.02\\
$a_2 $  
& 1.40 $\pm$ 0.05     
& 1.34$\pm$0.1 
& 0.915$\pm$0.005 
& 0.438$\pm$0.015 \\
$a_3 $  
& $-$1.00 $\pm$ 0.06  
& $-$1.15$\pm$0.1 
& $-$0.85 (input)
& $-$0.70$\pm$0.04 \\
$a_5 $  
& 3.30$\pm$0.05       
&  3.1$\pm$0.5 
&  2.700$\pm$0.001
&  1.29$\pm$0.04\\
$\tilde b_1 + \tilde b_2 $    
& 2.40$\pm$0.3   
& 2.60$\pm$0.2 
& 3.95$\pm$0.04
& 3.06$\pm$0.3  \\
$\tilde b_3 $                 
&$-$2.8$\pm$0.6  
& $-$3.96$\pm$0.9
& -1.45$\pm$0.03
& $-$0.41$\pm$0.27   \\
$\tilde b_6 $                 
& 1.4$\pm$0.3    
& 0.55$\pm$0.5 
& $-$1.17$\pm$0.17
& $-$1.5$\pm$0.2  \\
$ b_{16}- \tilde b_{15} $     
& 6.1$\pm$0.6    
& 7.1 $\pm$0.5 
&5.86 $\pm$0.19
&7.4 $\pm$0.5     \\
$ b_{19}$                     
& $-$2.4$\pm$0.4 
& $-$1.72$\pm$0.3 
& -0.44$\pm$0.21
&$-$3.7$\pm$0.2
   \\
\end{tabular}
\end{center}
\vspace*{-.3cm}
\caption[pepe]{\footnotesize HBChPT low energy constants.
Those in the first column were obtained from
fitting the {\it extrapolated} threshold 
parameters $a_{0,+}^\pm $, $b_{0,+}^\pm $, 
$a_{1,\pm}^\pm $, the Goldberger-Treiman discrepancy and the
nucleon $\sigma-$term to the HBChPT predictions~\cite{Mo98}.
In the second column we give the parameters
obtained from a low energy fit to $\pi N$ phase-shift data \cite{FMS98}.
In the third column we give the parameters obtained from an IAM fit 
with the $a_i$ constrained to the ranges predicted by resonance saturation.
Finally, the fourth column is the result of a totally unconstrained
fit.}
\end{table}
\begin{table}
\vspace{-0.5cm}
\begin{center}
\begin{tabular}{||c|c|c|c|c||c||c|c||}
  &  $1/F^2$  & $1/(F^2M) $ & $ 1/(F^2M^2)$ & $1/F^4$ & Total & 
SP98  & KA85\\
\hline
$a_{3\,1}^0 $  
&  $-$0.65        & $-$0.04         &  +0.07       & $+$0.07    
& $-$0.55$\er{0.16}{0.18}$  & $-$0.64 $\pm$ 0.01 & $-$0.72  \\
$a_{1,1}^0 $  
&  1.3       & $-$0.34          &  $-$0.08        &  0.07    
&  0.94$\pm$0.23  & 1.27 $\pm$ 0.02 & 1.26 \\
$a_{3\,3}^1 $  
&  35.3       & 47.95          & $-$1.75        & 0.26    
& 81.8$\er{0.8}{0.9}$  &  80.3$\pm$0.6  &  78.75 \\
$a_{1\,3}^1 $  
&  $-$17.7       & 15.46          & $-$3.1        & $-$6.35    
& $-$11.66$\pm 0.9$  & $-$10.5 $\pm$ 0.9 & $-$11.00\\
$a_{3\,1}^1 $  
&  $-$17.7       & 12.96          & $-$1.84        & $-$9.77    
& $-$16.3$\er{1.0}{0.9}$  & $-$15.9 $\pm$ 1.0 & $-$16.19\\
$a_{1\,1}^1 $  
&  $-$70.7       & 85.6          & $-$3.77        & $-$46.13    
& $-$35.0$\er{1.6}{1.5}$  &  $-$27.0 $\pm$1.4 & $-$28.67  \\
\end{tabular}
\end{center}
\vspace*{-.3cm}
\caption[pepe]{\footnotesize HBChPT, lowest partial S and P wave- scattering 
lengths, $a_{2I\,2J}^l $ (in ${\rm GeV}^{-2l-1} $ units) for $\pi N $ 
scattering using set II in Table I, 
decomposed as a sum of terms of first order 
$ 1/F^2 $, second order, $ 1/(F^2 M ) $, and third order, $ 1/(F^2 M^2)$ and 
$ 1/(F^4) $. The sum of all the terms yields the total scattering length 
parameter. For brevity, we only quote the errors in the final sum. 
The experimental values come from Ref.~\cite{AS95} for SP98,
and Ref.~\cite{K86} for KA85. For very recent and accurate
values of only the S-wave scattering lengths see 
\cite{nuevos}}
\end{table}
%
The IAM is based on the fact that elastic unitarity imposes 
the following relation
on a
generic partial wave $f$ and its inverse $f^{-1}$ (we drop the
$l$, $I$ and $J$ labels for simplicity):
\begin{equation}
{\rm Im}\, f = q\, |f|^2 \quad \Rightarrow \quad 
{\rm Im}\, f^{-1} = - q
\end{equation}
As a consequence, any amplitude satisfying exactly  elastic 
unitarity has the form
\begin{equation}
f=\frac{1}{{\rm Re}\, f^{-1}-i\,q}
\label{generalIAM}
\end{equation}
Thus, we only have to calculate ${\rm Re}\, f^{-1}$, whose different
approximations provide different
unitarization methods. 
Here we consider its expansion in terms of $m^2 /F^2 $, i.e.
\begin{equation}
f (\omega, m , F , M) 
= {m\over F^2}   t^{(1)} ( \omega/m \, , \, m/M) +
                     {m^3\over F^4}   t^{(3)} ( \omega/m \, , \, m/M) + \dots
\end{equation}
where we have used that $t^{(2n+1)}$ are dimensionless
functions, 
only depending on dimensionless variables. 
The functions $t^{(1)}$ and $t^{(3)}$ are only known in a further $m/M$ 
expansion,
\begin{equation}
t^{(2n+1)} ( \omega/ m  , m/ M) =  t^{(2n+1,2n+1)}
(\omega/m) + {m\over M} \,
t^{(2n+1,2n+2)} (\omega/m) + \, 
\left({m\over M}\right)^2 \, t^{(2n+1,2n+3)} 
(\omega/m) \, \dots
\end{equation}
yielding Eq.~(\ref{eq:fexpansion}) and Eq.~(\ref{eq:fexpansion2}) after 
a suitable isospin projection\footnote{ The relation between $ f_{2I,2J}^l $ 
and $ f_{l,\pm}^\pm $ is given by  $ f_{3,2l \pm 1}^l 
= f^+_{l,\pm} - f^-_{l,\pm}  $ and  $ f_{1,2l \pm 1}^l = f^+_{l,\pm} 
+2 f^-_{l,\pm}  $. }.

Perturbative unitarity in this expansion requires, 
\bea
{\rm Im}\, t^{(1)} = 0 \quad ,\quad{\rm Im}\, t^{(3)} = {q\over m} 
|t^{(1)}|^2,
\label{pertunit}
\eea
which also imply an infinite conditions in the $1/M$ expansion. 
 For $f^{-1}$ we get then 
\be
{1\over f } =
{F^2 \over m } {1\over t^{(1)}}
 -m { t^{(3)} \over
[ t^{(1)}]^2 } + \dots
\label{eq:invf2}
\ee
Obviously, expanding in $m/F^2$ may be justified provided these corrections
are small. As we have shown, for the $P_{33}$ channel, they are 
small precisely  at threshold, and there the unitarization 
scheme will, approximately, reproduce the perturbative result. At the same 
time, unitarity is exactly implemented since, thanks to eqs.(\ref{pertunit})
the above formula is of the form given in 
eq.(\ref{generalIAM}). 
However, the $1/M$ terms are not 
small corrections at threshold, and thus we do not further 
``expand the denominator''. In this way we keep the first mass 
corrections as equally important. At present, only the $t^{(1,1)}$, 
$ t^{(1,2)}$, $ t^{(1,3)}$ and $ t^{(3,3)}$ HBChPT terms are 
known (see our previous footnote \ref{foot}).
Therefore, we do not know neither $ t^{(3,4)}$ nor
$ t^{(3,5)}$. i.e $ O(1/(F^4 M))$ and $O(1/(F^4 M^2))$ respectively,
and we can only approximate the numerator of the second term by 
$t^{(3)}\simeq t^{(3,3)}$. That is, we know the first three orders of
the denominator $1/m$ expansion, but not  their 
counterparts in the numerator. Therefore, although 
we could use $t^{(1)}\simeq t^{(1,1)}+t^{(1,2)}+t^{(1,3)}$ in
the denominator,
for consistency with the numerator expansion we keep only
$t^{(1)}\simeq t^{(1,1)}$.
Incidentally, this corresponds to the static limit of the second term.
Note also that strict unitarity is still satisfied.
Of course, it would be  desirable to compute at least $ t^{(3,4)}$ 
and $ t^{(3,5)}$ 
in order to be able to keep also $ t^{(1,2)}$ in the denominator of the 
second term of the mentioned equation. 
After these remarks, we have  
\begin{equation}
{1\over f|_{\rm Unitarized}} =
{F^2 \over m } {1\over t^{(1,1)} + {m\over M} \,
t^{(1,2)} + \, ({m\over M})^2 \, t^{(1,3)} } 
 -m { t^{(3,3)} \over
[ t^{(1,1)} ]^2   } \label{eq:erajniam} 
\end{equation}

At threshold, our formula yields a modified  scattering length 
\bea 
{1\over a_{\rm Unitarized}} &=&
{F^2 \over m } {1\over \alpha^{(1,1)} + {m\over M} \,
\alpha^{(1,2)} + \, ({m\over M})^2 \, \alpha^{(1,3)}  }  
 -m { \alpha^{(3,3)}  \over
[ \alpha^{(1,1)} ]^2   } \\
&=& {1\over a_{\rm HBChPT} -({m^3 \over f^4})\,  \alpha^{(3,3)} }  
- m { \alpha^{(3,3)}  \over [ \alpha^{(1,1)} ]^2   } 
\eea 
which, using set II in Table I, yields 
$ a_{3 \,3}^1|_{\rm Unitarized}  = 82.95 \er{0.34}{0.35} {\rm
  GeV}^{-3} $, 
to be compared with 
$ a_{3 \,3}^1|_{\rm HBChPT}  = 81.8 \er{0.8}{0.9} {\rm GeV}^{-3} $, 
both compatible  with the experimental values. 
  Notice that if
we {\it had} expanded the denominator considering the second 
order contribution $O(1/(F^2M))$  to be small we  would have obtained 
strictly the IAM method as used in Ref.~\cite{GP99},  
\be  
{1\over a_{\rm IAM}} =
{F^2 \over m } \left\{ {1\over \alpha^{(1,1)}}
- {m\over M}\, { \alpha^{(1,2)} \over [ \alpha^{(1,1)}]^2 }
- \left({m\over M}\right)^2 \, { \alpha^{(1,3)} \over [\alpha^{(1,1)}]^2 }
+ \left({m\over M}\right)^2 \, { [\alpha^{(1,2)}]^2 \over [\alpha^{(1,1)}]^3 } \right\}
 -m { \alpha^{(3,3)}  \over  [ \alpha^{(1,1)} ]^2   } 
\ee
yielding $a_{3 \,3}^1 |_{\rm IAM}= 22.8 {\rm GeV}^{-3} $. 
This explains why an unconstrained IAM fit
with standard HBChPT leads to LEC which
are so different from those found in \cite{Mo98}, as noted in
Ref.~\cite{GP99}.

\section{ Numerical Results}

From our previous discussion it is clear that at threshold our unitarized 
amplitude  will reproduce very accurately and within error bars the HBChPT 
results and hence the experimental data. In addition, we expect that the 
$P_{33}$ phase shift can be extended up to the resonance region 
by propagating the errors of the LEC. It is also tempting
to extend the other five $S$ and $P$ wave phase shifts.
We show in Fig.1 the phase 
shifts obtained from our unitarization method, Eq.~(\ref{eq:erajniam}), 
compared with the experimental $\pi N $ data \cite{AS95}. 
The shaded area corresponds to the phase shifts obtained by propagating 
the errors of the parameters given in Ref.\cite{FMS98} (See Table I, set II)
by means of a Monte Carlo
gaussian sampling of the LEC  for any given CM energy
value.
Only for comparison, the dotted line corresponds to 
standard HBChPT extrapolated to high energies.

As one can see from the figures, the prediction of our unitarized
approach produces a distinctive resonance in the $P_{33}$ channel,
with very similar parameters to the physical $\Delta$ as we will see below.
Concerning the other channels, there is some improvement in the $S$
waves,
and a worse behavior for the $P_{13}$, $P_{31}$ and $P_{11}$, but 
note that these three partial waves have very tiny phase shifts, and any 
small error yields a large relative deviation.

The mass and width of the $\Delta$ resonance 
can be obtained either from the phase shifts,
by means of $\delta^1_{33}|_{s=M_\Delta^2}=\pi/2$ and
$1/\Gamma_\Delta=M_\Delta (d\delta^1_{33}/ds)\vert_{s=M^2_\Delta}$,
or from its associated pole  in the 
\begin{figure}[htbp]
\begin{center}                                                                
\leavevmode
\epsfysize = 400pt
\makebox[0cm]{\epsfbox{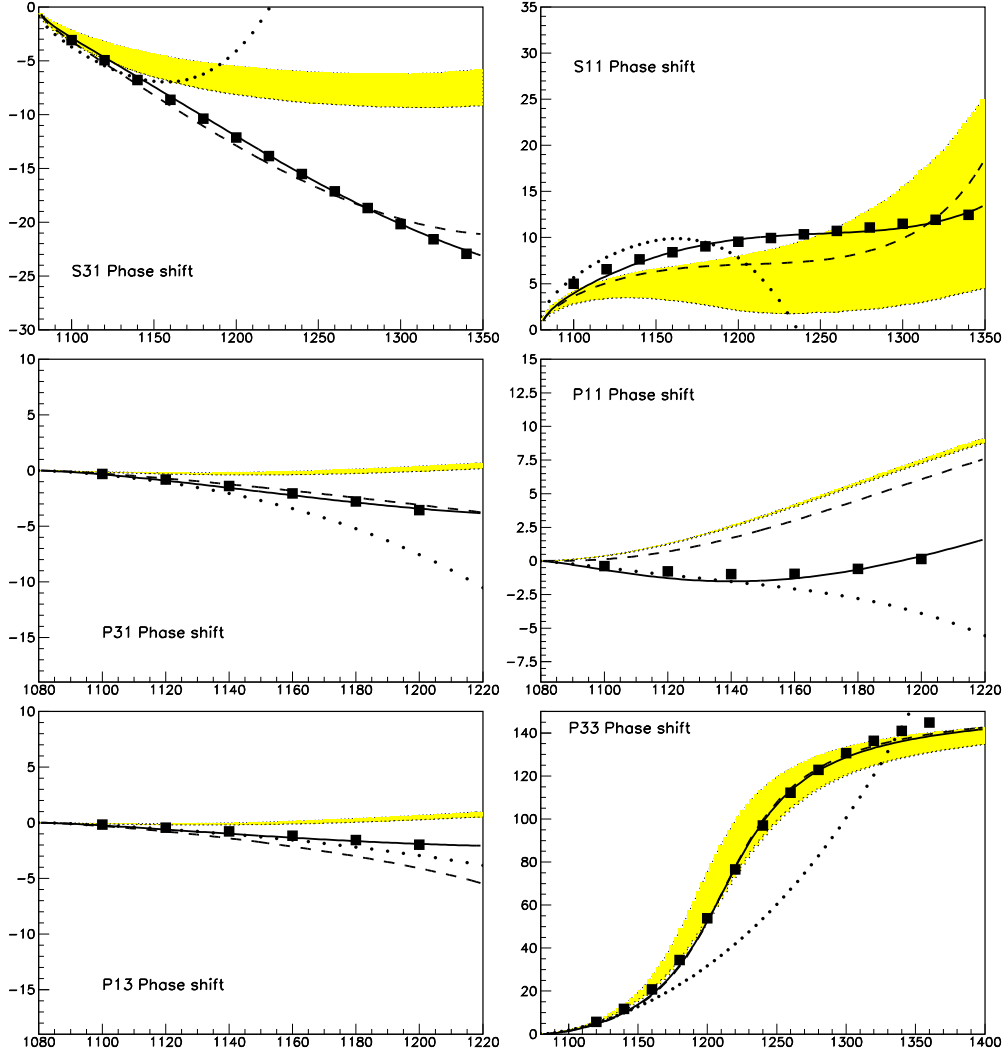}}
\end{center}
\caption[pepe] {\footnotesize Phase shifts as a function of
the total CM energy $\protect\sqrt s$. Experimental data are from
Ref.~\cite{AS95}. The shaded areas correspond to the propagated errors of the 
parameters in set II, which were obtained from low-energy data. The dotted line
is the HBChPT result extrapolated to high energies. The dashed line
is a fit with our unitarization procedure constrained to
Resonance Saturation, whereas the continuous line corresponds to 
an unconstrained fit.}
\end{figure}
\noindent second Riemann sheet 
($\sqrt{s_{pole}}\simeq M_\Delta - i\Gamma_\Delta/2$). 
We give in Table III the results
for different parameter sets. Note that the width of the ``predicted'' 
resonance from the parameters determined from low-energy data (set II) is
qualitatively very similar to the real $\Delta$. Of course, once
we fit to the data (set III and IV), 
we obtain a much better description.

 However, it was pointed out in Ref.\cite{GP99}
that the direct fit using the IAM directly on the HBChPT series
leads to chiral parameters of very unnatural size. 
That is not the case when we fit with the reordered method proposed here
as it can be seen in set IV of table I.
This set comes from an unconstrained fit to the six $S$ and $P$
wave $\pi N$ phase shifts, which is represented as
a continuous line in Fig.1.
For the fits, which start at 1130 MeV, 
we have used the MINUIT minimization routine assigning 
a 3\% uncertainty as in \cite{FMS98} plus a systematic error of one degree
to the data in \cite{AS95} (similar treatments are followed 
in \cite{MO99,GP99}).

Much more interesting \cite{GP99} are those fits where the $O(p^2)$ 
parameters are
constrained to the range estimated by the Resonance Saturation 
Hypothesis \cite{BKM97}. The fitted parameters are given as 
set III in Table I,
and the result is represented as the dashed line in Fig.1.
In this case the $P_{11}$ and $S_{11}$ are not so well described,
which may be due to effects of the $N^*(1440)$ and $N(1535)$, respectively,
which are the closest resonances to the energy regions displayed in Fig.1.
Indeed, the former plays a marginal role in the Resonance Saturation Hypothesis
whereas the latter is not even considered. 

A particularly relevant feature of these fits is that 
not only the resulting parameters have a more 
natural size, but also the $\chi^2$ per d.o.f. is between three and four
times smaller than for the corresponding
IAM fit applied to the standard HBChPT ordering. From this we can conclude
that considering the $1/M$ expansion separately as in our formalism 
is a sensible approach, apart from the details of its precise realization.

\begin{figure}[htbp]
\begin{center}                                                                
\leavevmode
\epsfysize = 200pt
\makebox[0cm]{\epsfbox{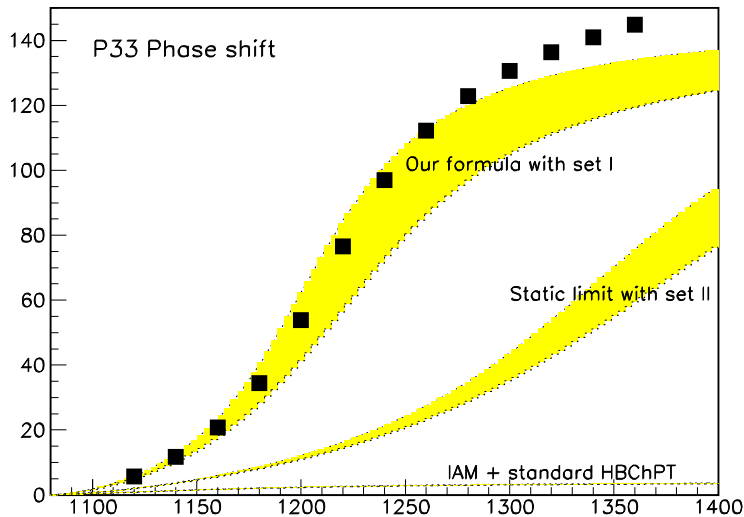}}
\end{center}
\caption[pepe]
{\footnotesize $P_{33}$ phase shift. The upper shaded
area corresponds to the result of propagating the errors of the 
set I parameters. This illustrates the uncertainties due to 
the choice of parameter sets from the literature. The intermediate
shaded area cover the propagated uncertainties of set II if
we applied the IAM in the static limit. Finally, with the same set
we show, in the lowest shaded area the result of the IAM when applied
directly to the standard HBChPT expansion.}
\end{figure}
\begin{table}[htbp]
  \begin{center}
    \begin{tabular}{|c|c|c|c|c|c|}
&set {\bf I}&set {\bf II}& set {\bf III} & set {\bf IV} & PDG \cite{PDG}
\\\hline\hline
\rule[-.3cm]{0cm}{.8cm}
$M_\Delta $  (MeV)& 1240 \er{23}{17} 
& 1222 \er{15}{12}    & 1227 $\pm 4$  
& 1226$\er{14}{12}$   & 1230 - 1234\\ \hline
\rule[-.3cm]{0cm}{.8cm}
$\Gamma_\Delta$ (MeV)&  157$\er{46}{33}$ 
&  117$\er{24}{18}$  &  104.1 $\er{5.2}{4.4}$ 
& 107$\er{21}{16}$ & 115-125 \\\hline\hline
\rule[-.3cm]{0cm}{.8cm}
Re (pole position) (MeV)& 1205
& 1204  
& 1204&  1204 & 1209-1211 \\ \hline
\rule[-.3cm]{0cm}{.8cm}
$\Gamma$ -2 Im (pole position) (MeV)&  110
&  110 &  110  
& 84 & 98-102 \\
    \end{tabular}
  \end{center}
\caption{$\Delta$ resonance parameters. The two first rows are
obtained from the condition $\delta^1_{33}|_{s=M_\Delta^2}=\pi/2$ and
$1/\Gamma_\Delta=M_\Delta (d\delta^1_{33}/ds)\vert_{s=M^2_\Delta}$.
The second two rows are the real part and minus twice the
imaginary part of its associated pole position,
calculated for the central values of each set 
(the errors are expected to be of the same order than for the
first two columns). }
\label{tab:deltapa}
\end{table}

As an illustration of the uncertainties due to the different
determinations of chiral parameters, we have also shown in Fig.2
the area corresponding to our formula applied to set I. We reobtain
the same qualitative result, although 
numerically the mass and width of the $\Delta$ are worse
than those obtained with set II.
In addition, in order to estimate the
convergence rate of our calculation, we have
also plotted in Fig.2 the prediction in the static limit ($ M
\to \infty $). The shaded area in Fig.2 corresponds to the 
propagated errors of the parameter set II in this limit.
As we see, there is also a distinctive resonant behavior, so that
the bulk of the dynamics is contained in the static limit.
However, the finite mass corrections, 
particularly the $1/(F^2 M) $ contribution, are
important to achieve a better description\footnote{The $1/(F^2
M^2) $ correction turns out to be quite small. That was 
expected, since it neither provides a sizeable contribution at
threshold (unlike the $1/(F^2 M)$ correction ) nor it is responsible
for the restoration of unitarity (like the $1/F^4$ correction).}.

We have also studied what happens if one includes the incomplete
higher order contributions to the second quotient in
Eq.~(\ref{eq:invf2}), i.e. if one approximates $t^{(1)}\simeq
t^{(1,1)}+t^{(1,2)}+t^{(1,3)}$ in its denominator. In such case we
obtain a worse result, closer to that of the static limit, but still
there is a distinctive resonant behavior, improving the IAM results
with the standard HBChPT expansion.  This suggests that the unknown
higher order contributions (see footnote \ref{foot}) in the numerator
could give rise to some cancellation with those still incomplete of
the denominator. Finally, we also show in Fig.2 the results obtained
within the conventional IAM approach, for the parameter set II.  As it
was already pointed out in \cite{GP99} the result is extremely poor if one
uses the parameters determined from low energy data.

\section{Conclusions and Outlook}

Heavy Baryon Chiral Perturbation Theory provides 
definite predictions for the $\pi N$ scattering amplitudes in the threshold 
region.  However it violates exact unitarity if the perturbative expansion 
is truncated to some finite order and also is unable to describe the 
$\Delta$ resonance (and its associated pole) in the $P_{33} $ channel. 
The analysis up to third order 
shows that the leading finite nucleon mass correction, which is second order, 
is of comparable size to the static approximation and in fact it dominates 
the corrections at threshold. This suggests a unitarization method
using the expansion in inverse powers of the weak pion decay constant
but without making the heavy baryon expansion. Such an idea is supported by 
recent theoretical attempts to redefine a relativistic chiral counting 
for baryons. We have proposed a unitarization scheme based on the Inverse
Amplitude Method applied to this reordered HBChPT expansion. 
It provides a {\it prediction} for the $\pi N$ phase 
shifts, which generates a $\Delta$ resonance 
from the low energy constants and their 
errors, as determined from  HBChPT. The fits within this scheme
provide chiral parameters of a natural size  and a better 
overall description than those performed with the IAM applied to
the HBChPT standard expansion.
This result suggests that including the $1/M$ expansion separately
is a sensible physical approach. In addition, this method can be easily
generalized to higher orders and coupled channels. Further work along these 
lines is in progress.

\section*{Acknowledgments}
This research was supported by DGES under contract PB98-1367 and by
the Junta de Andaluc\'\i a. Work partially supported by DGICYT under 
contracts AEN97-1693 and PB98-0782. 
We thank P. B\"{u}ttiker, J.A.Oller and E.Oset for useful comments and 
discussions.

\end{document}